\documentclass[twocolumn,amssymb,aps,prl,longbibliography]{revtex4-2}
\usepackage{amsfonts}
\usepackage[utf8]{inputenc}
\usepackage[english]{babel}
\usepackage{amsmath}
\usepackage{graphicx}
\usepackage{color}
\usepackage{amsmath}
\usepackage{bm}
\usepackage{xcolor}
\usepackage{array}
\usepackage{multirow}
\usepackage[colorlinks=false, linkbordercolor=red, citebordercolor=green, urlbordercolor=cyan, pdfborderstyle={/S/U/W 1}]{hyperref}
\usepackage{comment}
\usepackage[normalem]{ulem}

\begin{document}

\title{Quantum non-Gaussian coherences of an oscillating atom}
\author{A.~Kovalenko$^{1,*}$, L.~Lachman$^{1,*}$, T.~Pham$^{2}$, K.~Singh$^{1}$, O.~\v{C}\'{i}p$^{2}$, L.~Slodi\v{c}ka$^{1}$, and R.~Filip$^1$}

\affiliation{$^1$ Department of Optics, Palack\'{y} University, 17. listopadu 12, 771 46 Olomouc, Czech Republic \\
$^2$ Institute of Scientific Instruments of the Czech Academy of Sciences, Kr\'{a}lovopolsk\'{a} 147, 612 64 Brno, Czech Republic}

\begin{abstract}
Quantum coherence between energy eigenstates of harmonic oscillators is essential for quantum physics. Even the most elementary binary superpositions of the ground and the higher eigenstate are highly required for quantum sensing, thermodynamics, and computing. We derive upper bounds for quantum coherences achieved by classical and Gaussian states and operations and, subsequently, obtain a hierarchy of the thresholds for the off-diagonal elements necessary to reach genuine quantum non-Gaussian coherences. We experimentally demonstrate unambiguous observation of quantum non-Gaussian coherences in mechanical vibrations of a single calcium ion up to the superposition of zero and six phonons. The analysis of the robustness with respect to pure dephasing in a motional Ramsey experiment demonstrates the feasibility of their storage for up to more than 20~ms for superpositions with a large energy difference of participating number states. The presented observations prove that atomic oscillations go deeply into a diverse area of discrete quantum non-Gaussian coherent phenomena critical for their applications.
\end{abstract}

\maketitle

\def\thefootnote{*}\footnotetext{These authors contributed equally to this work}

\section{Introduction}

Coherence is an essential notion of modern science. Fundamental discoveries of the microwave maser, optical laser and Bose-Einstein condensation of cold atoms have largely boosted the classical coherence of radiation and matter~\cite{mandel1965coherence,cronin2009optics}. Modern quantum mechanics, thermodynamics and applications of the second quantum revolution additionally use essential quantum coherences~\cite{streltsov2017colloquium}. Differently from classical, they are fundamentally present in the oscillators' superposition of energy eigenstates. However, classical external drive is only capable of Gaussian quantum coherences on linearized oscillators. They are sufficient for basic quantum sensing~\cite{giovannetti2004quantum,pezze2018quantum} and point-to-point continuous-variable quantum communication~\cite{pirandola2020advances}, but insufficient for many advanced applications~\cite{eisert2002distilling,niset2009no,broz2023test}, including quantum computation~\cite{lloyd1999quantum}.
A pioneering example is a superposition $(|0\rangle+|n\rangle)/\sqrt{2}$ of the oscillator ground and higher energy eigenstate (0N~states). They are already experimentally applied to reduce estimation errors in frequency sensing~\cite{mccormick2019quantum,wang2019heisenberg} or construct binomial quantum error correction codes~\cite{hu2019quantum}, recently approaching the breakeven point. Such superpositions may further upgrade bosonic quantum communication~\cite{burkhart2021error}, extend quantum engines~\cite{maslennikov2019quantum}, advance quantum simulations of chemical processes~\cite{wang2020efficient}, or witness gravitation effects~\cite{howl2021non}.

Achieving these milestones by quantum non-Gaussian nature of the Fock-state superpositions depends on the off-diagonal elements of the density matrix. Despite pioneering tests of their applications with visible quantum non-Gaussian nature of the overall state, quantum coherence present in such off-diagonal elements has not been explicitly and conclusively proven to be quantum non-Gaussian. It is essential to claim supremacy by quantum non-Gaussian coherences in both the fundamental tests and applications. Quantifying non-Gaussianity in diagonal elements of the density matrix in energy (Fock) states is insufficient, as the off-diagonal elements carry additional quantum resources. Moreover, the measurement of off-diagonal components is conceptually different, requiring an interferometric setup~\cite{ramsey1949new,maitre1997quantum,wineland1998experimental,cronin2009optics}. In contrast, the diagonal elements can be measured by projecting and evaluating the Fock states $|n\rangle$~\cite{podhora2022quantum}. Unfortunately, the fidelity with a superposition in the 0N state obtained by projection~\cite{chabaud2021certification} undesirably averages information from diagonal and off-diagonal elements. Superpositions of two Fock states will comprise two relevant and independent quantum non-Gaussian ranks, one related to the Fock state and new one related to the quantum coherence purely specified by the off-diagonal element. Diverse natural couplings to the thermal environment will affect the feasibility of their observation differently. For example, dephasing applied to such a state will affect exclusively the rank related to the coherence.
Therefore, ideally a hierarchy of the faithful criteria for quantum non-Gaussian (QNG) coherences for individual off-diagonal components capable of proving increasingly powerful quantum coherent resources is required. For a long time, it has been missing, and experimental conditions to conclusively observe such intriguing QNG coherences have been unclear and challenging.

Here, we derive the hierarchy and experimentally demonstrate quantum non-Gaussian coherences of 0N states up to the superposition of the Fock states $|0\rangle$ and $|6\rangle$ with a single trapped ion. They indicate building up genuine QNG coherences, distinguishing them from these still partially provided by Gaussian tools of linearized dynamics, for example, in the case of displaced Fock states~\cite{mccormick2019coherently}. Analysis of their depth against pure dephasing allows physical comparison of the experimental results. For example, the 0N state with $n=4$ relevant for binomial codes~\cite{michael2016new} has a twice as high QNG coherence depth than a higher 0N code state with $n=6$. We compare such QNG features also with other binary Fock-state superpositions. Finally, we observe a storage of the depth of the QNG motional coherences of a single calcium ion, surviving for more than 20~ms for the binomial code state. Our local hierarchical approach to quantum non-Gaussian coherence refines the global coherence measures based on $l_1$ norm~\cite{streltsov2017colloquium}.

\section{Preparation and measurement of quantum coherence}

A motion of atomic ions trapped and laser-cooled in a linear Paul trap provides a unique test bed for the pilot observations of a QNG coherence. The combination of feasibility of high motional frequencies on the order of a few MHz, exceptional trapping potential stability, and low heating rates reaching less than a single phonon per second achievable in modern bulk 3D linear traps, has already resulted in a number of notable experimental tests in which the mechanical QNG properties of low Fock states and even their motional coherent superpositions corresponded to the paramount resources~\cite{wolf2019motional,mccormick2019quantum,fluhmann2019encoding}. As illustrated in Fig.~1-a), a single trapped ion oscillating in the harmonic pseudopotential at frequency $\omega_m$ is akin to a very pure linear mechanical pendulum with a well-defined discrete internal electronic level structure and with the feasibility of the unprecedented precise control in the linear and non-linear regime~\cite{leibfried2003quantum,akerman2010single}. Its mechanical properties are resulting from the Coulomb forces implemented through the trapping electrodes and from interaction with laser beams. While thermalization and heating caused by the residual coupling to the surrounding environment are intrinsic to a majority of feasible quantum mechanical oscillator platforms~\cite{aspelmeyer2014cavity}, atomic ions in Paul traps allow for a very unique combination of extremely low heating rates and deterministic nonlinear interactions in the quantum regime~\cite{roos2008nonlinear,ding2018quantum} even in room temperature setups.
The laser interaction with the electronic degrees of freedom of an ion provides crucial deterministic and highly precise coupling to the motional degrees of freedom, which can be addressed individually and with the feasibility of highly nonlinear interactions~\cite{leibfried2003quantum}.

A coherent superposition of motional Fock states can be achieved deterministically solely by interrogating the resonant laser interactions on the carrier, first blue (BSB), and red (RSB) motional sidebands~\cite{meekhof1996generation,mccormick2019quantum,gardiner1997nonclassical}. The presented experimental test employs an axial motion at $\omega_{\rm m}=(2 \pi)\times 1.11$~MHz of a single trapped $^{40}{\rm Ca}^+$ ion in a linear Paul trap and its electronic quadrupolar transition $|g\rangle:=4^{2}{\rm S}_{1/2}(m=-1/2)\leftrightarrow |e\rangle:=3^{2}{\rm D}_{5/2}(m=-1/2)$. Despite the high level of mechanical control of a trapped ion, the thermalization of its motional amplitude through coupling to surrounding thermal noise can’t be fully suppressed~\cite{brownnutt2015ion}. Its impact on the ion cooled close to the motional ground state can be well characterized by quantifying the rate of the motional heating, which was found $3.2\pm 0.2$~phonons/s for the given axial motional frequency. We note that this heating sets the fundamental upper limit on the motional coherence. The decoherence timescale observable on the electronic $|g\rangle \leftrightarrow |e\rangle$ transition estimated as to be much shorter, about 8~ms. Therefore, the target motional superpositions are realized such that they are decoupled from the electronic transition.

Fig.~1-b) illustrates the Ramsey interferometry on mechanical 0N superpositions of a single ion~\cite{mccormick2019quantum,jarlaud2020coherence,milne2021quantum}. A state initialization corresponding to laser cooling and optical pumping steps prepares a spin-motional ground state $|g,0\rangle$ with a high probability of more than~0.98. A BSB laser pulse with an area of $\pi/2$ realizes deterministic coherent splitting of motional and electronic populations resulting in spin-motional entanglement $(|0,g\rangle + \exp(-i \phi_1)|1,e\rangle)/\sqrt{2}$, where $\phi_1$ is the phase of the first optical pulse in the Ramsey sequence. The following composite deterministic population transfer results in states with targeted solely motional superpositions. It corresponds to the application of the RSB and BSB $\pi$-pulses which transfer the motional population from $|1,e\rangle$ through the combined qubit-oscillator state ladder. The increase of the phonon number is accompanied with coherent flips of the electronic state per each laser pulse and accumulation of the overall superposition phase forming $\phi_{\rm R_1}=\sum_{l=1}^{k} \phi_l$, where $\phi_l$ is the phase resulting from the interaction with $l$-th pulse and includes accumulated phase shift in the coherent evolutions of intermediate superpositions of electronic and motional states. The whole preparation sequence corresponds to \textit{effective} $\pi/2$ composite pulse resulting in the state approaching the 0N superposition
\begin{equation}
|\psi_{0,n}\rangle=\frac{1}{\sqrt{2}}(|0\rangle+ \exp(i\phi_{\rm R_1}) |n\rangle).
\label{ONdef}
\end{equation}
For the $|0,n\rangle$ states with $n>2$, the preparation additionally includes the coherent optical \textit{shelving} of $|0,g\rangle$ population to the auxiliary $|a\rangle:=3^{2}{\rm D}_{5/2}(m=-5/2)$ level. This allows for independent deterministic manipulation of the excited $|1,e\rangle$ state population in the initial spin-motional superposition.

\begin{figure*}[!t]
\begin{center}
\includegraphics[width=2\columnwidth]{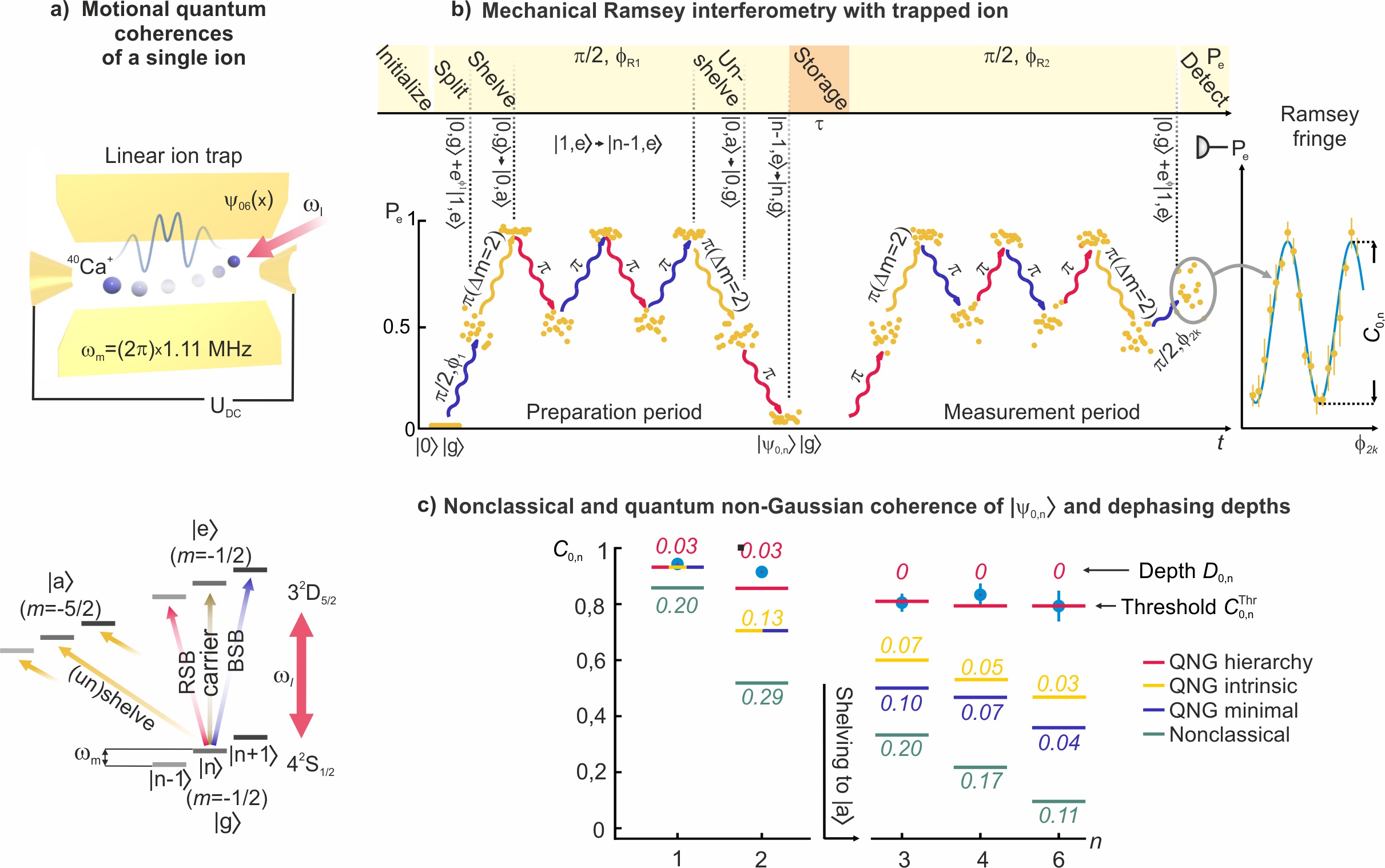}
\caption{The experimental observation of quantum non-Gaussian coherences on single atom mechanics. Part~a) depicts fundamental principles of the presented implementation of a coherently controllable mechanical object in a quantum regime. A single trapped $^{40}\rm{Ca}^+$ ion trapped and laser-cooled in a linear Paul trap provides crucial ingredients for achieving the QNG regime through coherent coupling to blue (BSB) and red (RSB) motional sidebands resulting from the modulation of the narrow optical transition $|g\rangle\leftrightarrow|e\rangle$ due to the axial motion with frequency $\omega_{\rm m}$. b) shows the example of the experimental pulse sequence implementing the effective motional Ramsey interferometer. The yellow points are the measured populations $P_{\rm e}$ after each sequence step. The crucial parts of the state generation correspond to the realization of a spin-motional superposition, shelving of the $|0\rangle|g\rangle$ state, and coherent transfer of $|1\rangle|e\rangle$ to higher motional number state $|n\rangle|e\rangle$. The overall pulse sequence corresponds to an effective $\pi/2$ pulse between motional number states $|0\rangle$ and $|n\rangle$ with a phase $\phi_{\rm R1}$. Following the controllable time delay~$\tau$, the second part of the interferometer realizes the inverse unitary with an overall phase $\phi_{\rm R1}$, which can be precisely scanned by the offset phase $\phi_{\rm k2}$ applied to the last BSB laser pulse. The sequence concludes by the estimation of the excited state probability $P_{\rm e}$ for the given controllable phase of the last analysis pulse $\phi_{\rm 2k}$. The repetition of the whole sequence provides the necessary suppression of the projection noise and results in the single data point in the interference fringe for estimation of the coherence amplitude $C_{0,n}(\tau)$.  The graph in~c) summarizes results for $|\psi_{0,n}\rangle$ superposition states at the zero Ramsey delay $\tau=0$ (without a storage), the corresponding fundamental theoretical thresholds $C_{0,n}^{\rm Thr}$ for nonclassical, quantum non-Gaussian, both minimal and intrinsic, and finally, hierarchy of quantum non-Gaussian coherence shown as horizontal lines. The listed numbers correspond to estimated mean values of experimental depths $D_{0,n}$ of these properties defined by~(\ref{depth}) gauged by a pure dephasing process. The measured step decrease in the coherence for the $|\psi_{0,n}\rangle$ states with $n>2$ is caused by the significantly greater complexity of the applied pulsed sequence, which additionally employs shelving of $|0,g\rangle$ to the auxiliary electronic state $|a\rangle$.}
\label{fig:block_scheme}
\end{center}
\end{figure*}

The coherence amplitudes are estimated using a Ramsey interferometry in the binary motional $|0\rangle, |n\rangle$ basis. The second composite \textit{effective} $\pi/2$ rotation is set to follow the inverse of the generation state preparation sequence and, therefore, closes the motional Ramsey interferometer~\cite{turchette2000decoherence,mccormick2019quantum,jarlaud2020coherence}. It maps the coherence amplitude of the $|\psi_{0,n}\rangle$  state on the amplitude of the off-diagonal matrix element of the final spin-motional superposition $(|0,g\rangle + \exp(-i \phi)|1,e\rangle)/\sqrt{2}$. Its accumulated phase $\phi$ is given by the sum of state preparation phase $\phi_{\rm R1}$, phase corresponding to the Ramsey delay given by $\phi_{\tau}=\delta\omega \tau$, and phase accumulated within the analysing Ramsey composite $\pi/2$ pulse $\phi_{\rm R2}=\sum_{l=k+1}^{2 k} \phi_{l}$. The phase of the last $\pi/2$ BSB analysis laser pulse $\phi_{\rm 2 k}$ is scanned to access the coherence of the final electronic state superposition propagating from the coherence of the generated $|\psi_{0,n}\rangle$ state.

A Ramsey waiting period with a duration~$\tau$ allows for a free precession of
the motional superposition and provides a tunable tool for the estimation of its temporal coherence stored in the mechanical oscillations.
Note that the phase accumulation due to the residual difference $\delta\omega$ between the free precession at the motional frequency and at the frequency of the reference oscillator is accelerated by a factor~$n$ proportional to the energy difference of the participating number states. A factor of the similar form is expected to contribute also during the state creation and analysis composite Ramsey pulses and corresponds to the stable residual offset between the axial motional frequency and the detuning of the sideband-excitation lasers from the carrier transition.

The sequence provides an estimation of the coherence amplitudes defined formally in the Fock state basis as
\begin{eqnarray}
C_{0,n}(t)= \max_{\phi} {\rm Tr} [X(\phi)\rho(t)]
-\min_{\phi}{\rm Tr} [X(\phi)\rho(t)]
\label{coherence}
\end{eqnarray}
where $X(\phi)$ is the measured observable in the Ramsey interferometer corresponding to on the off-diagonal element with phase $\phi$
\begin{equation}
X(\phi)=1/2 (1+[|0\rangle\langle n|\exp(i\phi)+|n\rangle\langle0|\exp(-i\phi)]).
\label{projector}
\end{equation}
The observation of the state-dependent fluorescence signal from the coupled $4^{2}{\rm S}_{1/2} \leftrightarrow 4^{2}{\rm P}_{1/2}$ electric dipole transition completes the projective measurement with a projector~(\ref{projector}) on state~(\ref{ONdef}) within an electronic state basis. The final probability of finding the electron in the excited state
\begin{equation}
P_{e}(\phi)=\frac{1}{2}(1+ C_{0,n}(\tau) \cos(\phi))
\label{Pe}
\end{equation}
then depends on relative Ramsey phases $\phi=\phi_{\rm R1}+\phi_{\tau}+\phi_{\rm R2}$ appearing in the motional interferometer. The off-diagonal elements $\langle n|\rho|m\rangle$ are a complex number, but as in any interferometry, we scan through an auxiliary phase $\phi_{2k}$ of the last pulse in the Ramsey sequence to find its largest real part $C_{0,n}(t)$. This off-diagonal element is not normalized in order to preserve the linearity of $C_{0,n}(t)$ in the state, which is important for the derivation of the criteria. Crucially, the estimated coherence amplitudes, as presented in Fig.~\ref{fig:block_scheme}-c) do not depend on the absolute value of the phase of the superposition and is estimated merely as the contrast of the observed motional Ramsey interference. It is thus not affected by any additional temporarily stable phase offsets emerging from practically unavoidable but stable phase evolutions, including for example periods between laser pulses. They manifest merely as a constant phase shift of the interference pattern. However, the overall phase $\phi$ must be stable on the timescales corresponding to the measurement of Ramsey interference fringes for the estimation of the single value of $C_{0,n}$. Such stability is achieved by targeting the motional states decoupled from an electronic transition and by the adjustment of the measurement duration with respect to the estimated drift rate of the axial motional frequency, as described in Supplementary Materials.

\section{Quantum non-Gaussian coherences}

The lowest threshold addresses nonclassical coherences, i.e. values of $C_{0,n}$ which are unfeasible for a convex set of coherent states. Coherent states can rise in a linear oscillator by linear classical external drive without needing nonlinearity. We observe the thresholds for an unambiguous proof of nonclassical coherences
to be favorably decreasing with the Poissonian tail as $n$ is larger, as presented in Fig.~\ref{fig:block_scheme}-c). The {\em minimal} QNG threshold $C_{0,n}^{\rm G,min}$ excludes all values of $C_{0,n}$ by a convex mixture of all Gaussian states~\cite{Yuen1976two} that can be obtained by maximally quadratic nonlinear oscillator dynamics.
However, in this narrowed set, the coherence can still come from a final Gaussian operation on the Fock states, we must extend the set of the rejected states further and refine our understanding of the QNG coherences. Therefore, we also exclude all mixtures of squeezed displaced Fock states, which results in thresholds $C_{0,n}^{\rm G, int} > C_{0,n}^{\rm G}$ on an \textit{intrinsic} QNG coherence. They bound values of off-diagonal matrix elements in Fock basis that can be reached by any Gaussian operations on mixtures of Fock states.

Still, the classification of quantum non-Gaussian coherence in the $C_{0,n}$ elements needs additional refinement. Squeezing and displacement can also act on some coherent superpositions up to Fock states $|n-1\rangle$ with ideal QNG coherences in a lower dimension and make coherence in $C_{0,n}$ by the Gaussian resource. This suggests the requirement of building and analysing a hierarchy of QNG coherences in analogy with the approach developed for the diagonal elements~\cite{lachman2019faithful,podhora2022quantum}. It quantifies the strong requirement that for any Fock-state superposition with up to to $n-1$ phonons, the step towards a coherence $C_{0,n}$ in superposition with $n$ phonons is not feasible with any combination of Gaussian squeezing and displacement. Mathematically, it excludes a convex closure of any Fock-state superposition $|\tilde{\psi}_{n-1}\rangle=\sum_{j=0}^{j=n-1} c_j |j\rangle$ up to $n-1$ followed by any Gaussian squeezing and displacement operation to build the coherence higher than the threshold $C_{0,n}^{\rm G_n}$, where $G_n$ denotes \textit{genuine} quantum non-Gaussian coherence of state $|\psi_{0,n}\rangle$. It extends the approach developed for the diagonal elements~\cite{lachman2019faithful,podhora2022quantum}, without mixing them in an overall state fidelity~\cite{chabaud2021certification}, allowing direct comparison of these distinctive genuine QNG aspects. The derivation of the corresponding thresholds on coherence amplitudes involves numerical optimizations presented in Supplementary materials. Differently than for diagonal elements~\cite{podhora2022quantum}, the thresholds decreases with $n$, as all the previous ones.

Fig.~\ref{fig:block_scheme}-c) presents a summary of the measured quantum non-Gaussian coherences from the Ramsey interference fringes for a zero Ramsey delay $\tau=0$ and for states approaching $|0,n\rangle$ and corresponding thresholds on the unambiguous identification of the nonclassical, quantum non-Gaussian and genuine quantum non-Gaussian character. The experimental coherences correspond to raw measured values without any correction on the known dephasing and experimental imperfections in the second part of the Ramsey interferometer. After the state preparation step, the coherence amplitudes are expected to be higher by nearly half of their residual to the ideal value of one. The experimentally observable coherences for $n=1$ and~2 are limited dominantly by the residual thermal energy after the laser sideband cooling estimated to $\overline{n}=0.07\pm 0.01$~thermal phonons and small imperfections in the applied coherent laser pulses including the off-resonant coupling to the radial motional modes. The upper limit on the target coherences is thus set by the state populations ${\rm max}(C_{0,n}) \approx 2\times \sqrt{p_0\times p_n}$, which explains measured data for states approaching $|\psi_{0,1}\rangle$ and $|\psi_{0,2}\rangle$ within experimental error bars. For $|\psi_{0,n}\rangle$ states with $n>2$, the additional step decrease of $C_{0,n}(\tau=0)$ by about 5\,\% resulted from inclusion of shelving and un-shelving process to the auxiliary level $|a\rangle:=3^2{\rm D}_{5/2}(m=-5/2)$. The corresponding quantum coherences thus still suffer from a step reduction due to additional dephasing of the internal-electronic states within the time periods spent in the electronic superpositions of different sub-levels of the $3^{2}{\rm D}_{5/2}$ manifold. The set of the four shelving laser pulses also limits the feasible coherence through accumulation of errors due to imperfect population transfer.

\section{QNG coherence depth under dephasing}

For comparison of QNG coherences between different experiments, we must use quantification that can be applied to any results despite the difference of such platforms. To evaluate the robustness of the observable QNG aspects, we employ the natural sensitivity of coherence to ideal dephasing until it does not fall below the lowest QNG threshold $C^{\rm G, min}_{0,n}$ that can be applied to interferometric measurements. The detected QNG coherences presented in Fig.~\ref{fig:block_scheme}-c) can be directly compared for various $n$ using the difference between data and corresponding thresholds, with the distance given by a pure dephasing, which will amount to its reduction according to
\begin{equation}
C_{0,n}(\tau+\Delta T)=\exp(-\frac{\Gamma \Delta T}{2} n^2) C_{0,n}(\tau).
\label{decayC}
\end{equation}
Here $\Gamma\Delta T$ corresponds to an effective variance of the Gaussian phase fluctuations in the coupling to thermal phase reservoir added to the respective coherence at given time $\tau$~\cite{myatt2000decoherence}. The depth of the observable QNG coherence at time~$\tau$ can then be defined by evaluating the amount of such \sout{coupling} dephasing $\Gamma\Delta T$ necessary for reaching the corresponding threshold values $C_{0,n}^{\rm G}$ (blue numbers in Fig.~\ref{fig:block_scheme}-c))
\begin{equation}
D_{0,n}(\tau)=\frac{2}{n^2}\ln \frac{C_{0,n}(\tau+\Delta T)}{C_{0,n}^{\rm G}}.
\label{depth}
\end{equation}

\begin{figure}[!t]
\begin{center}
\includegraphics[width=1\columnwidth]{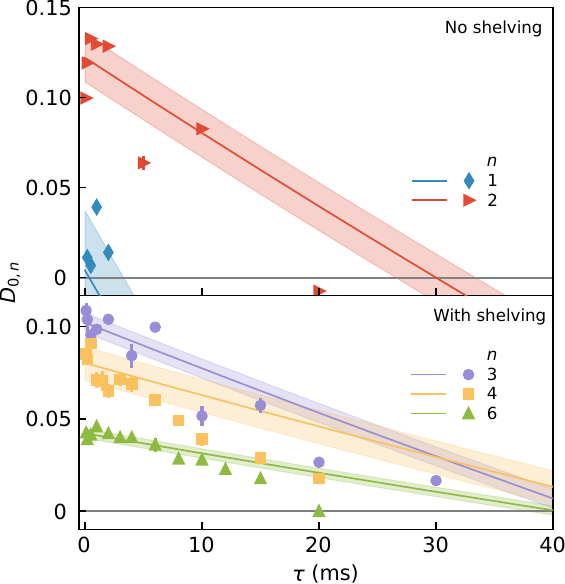}
\caption{The QNG depths $D$ according to Eq.~\ref{depth} for experimentally realized $|\psi_{0,n}\rangle$ superposition states for different Ramsey delay times provide a perspective on the natural loss of quantum coherence in a trapped ion mechanical oscillator. The initial values at $\tau\approx 0$ experience a rapid enhancement for $n=2$ due to the decrease of the corresponding thresholds $C^{\rm G, min}_{0,n}$ followed by the gradual decrease caused by the increased complexity and length of the pulse sequences for $n\geq 3$. The evident enhanced durability of QNG coherence in $|\psi_{0,n}\rangle$ with higher $n$ prevails despite the increase in sensitivity to thermalization of diagonal elements through inevitable heating of an ion in the trap, which sets the upper limit on the QNG depth achievable without any dephasing depicted with solid curves. The statistical uncertainty of the initial values $D(\tau \approx 0)$ corresponds to the dominant part of the uncertainty of this limit and is depicted as a colored area. The error bars correspond to an evaluated single standard deviation in both simulation and data. The rapid decrease of the variance of presented data points for higher $n$ is the consequence of $n^{-2}$ scaling of the QNG depth for $|\psi_{0,n}\rangle$ states in~(\ref{depth}).}
\label{decay}
\end{center}
\end{figure}

The theoretical and experimental depths of QNG coherences represented by the off-diagonal elements can be directly compared with depths of fidelities~\cite{chabaud2021certification}; see Table~1 in Supplementary Material. The resulting systematically larger depths of fidelities in both theoretical states and their experimental counterparts suggest that genuine QNG character of coherences should be evaluated separately from the diagonal elements. Fig.~\ref{decay} presents measurements of the temporal decay of the mechanical QNG coherence depth on motional states approaching $|0,n\rangle$ superpositions. The initial values of QNG depth $D_{0,n}(\tau\approx 0)$ acknowledge the expected increased difficulty to realize the quantum non-Gaussian coherence for motional superpositions with higher $n$ due to the increased complexity of the laser pulse sequence employing shelving method and, at the same time, the relative increase in the QNG depth for superposition states with higher~$n$. It overcomes the $n^{-2}$ scaling in~(\ref{depth}) and reflects the difficulty to approach high coherence for $|\psi_{0,n}\rangle$ states with large $n$ from any mixture of Gaussian states. Their nontrivial decay behaves in stark contrast to the conventional expectations based on Eq.~(\ref{decayC}). The robustness to decoherence beyond the QNG thresholds increases with $n$ as the corresponding slope of the measured depths decreases, which became obvious from the measured data despite the presence of the thermalization effect rising with~$n$. Generally, two different sources of coupling to an environment are expected to affect the observable temporal behavior of the
motional coherence of a trapped ion. They can be conceptually understood through the coupling of the motion to
the thermal amplitude or phase reservoirs~\cite{turchette2000decoherence}. The thermalization of the presented states is determined by the measured motional heating of the ion in the trap~\cite{brownnutt2015ion}. It limits the coherence under the constraints on the positivity of the density matrix, and thus represents an upper limit on $D$ achievable by fully suppressing the dephasing present in the experiment. However, we note, that the observed increase in the storage times for $|\psi_{0,n}\rangle$ with larger $n$ is expected also for the bare thermalization limit on the QNG depth, which is evident from the gradients of these thermalization limits.

Contrary to the closely related experimental efforts in optical or micro-wave oscillator platforms, where the coherence is dominantly limited by photon loss in the state manipulation and detection process~\cite{ourjoumtsev2007generation,vlastakis2013deterministically}, trapped ion oscillator decoherence is typically significantly impacted by thermalization and heating~\cite{turchette2000decoherence,milne2021quantum}. This emphasizes the similarity to the intrinsically mechanical systems, where the rates of phonon loss or dephasing can be suppressed much below the rate of thermalization due to the feasibility of mechanical designs with very high quality factors~\cite{delic2020levitated,tsaturyan2017ultracoherent,aspelmeyer2014cavity}.
On the other hand, in the regime of presented very low motional heating rates,
the observable storage is noticeably impacted at longer storage times by pure dephasing, which appears to be an important and previously mostly unexplored regime~\cite{turchette2000decoherence,milne2021quantum,brownnutt2015ion,mccormick2019quantum}.
The additional estimation of diagonal matrix elements for the data presented in~Fig.~\ref{decay} confirmed that their temporal evolution corresponds well to thermalization dynamics at the measured heating rate. It complementarily confirms that the presented residuals of the estimated coherence correspond solely to dephasing effects. The corresponding data of the measured coherence decay and estimation of their decay due to the coupling to the thermal amplitude reservoir characterized by the independent measurement of the motional heating rate can be found in Supplementary Materials.

The expected phase-damping source of motional coherence in this system corresponds to long-term fluctuations of the axial DC electric potential realized by the precise high voltage source. We estimated the corresponding average drift in the motional frequency to~$0.44 \pm 0.11$~Hz/min. In addition, noise spectral features in a~kHz domain could arise from electromagnetic pickup in the trap, which is expected to be significant in the decreased coherence amplitudes on the presented timescales~\cite{milne2021quantum}. However, the reliable precise estimation of the full motional noise spectrum is extremely challenging, even when employing recently proposed and implemented quantum sensing techniques~\cite{milne2021quantum,mccormick2019quantum,mccormick2019coherently}. We foresee a potential applicability of the presented states possessing provable QNG coherence for such efficient sensitive probing of motional spectra.

\section{Quantum coherence of $\mathbf{|\psi_{m,n}\rangle}$ states}

QNG coherence in a binary quantum superposition in the Fock state basis can behave very diversely. In comparison with $|\psi_{0,n}\rangle$ states, it might be expected that the opposite case of the superposition of $n-1$ and $n$ phonons become less sensitive to dephasing for $n>1$ and thus can be easier to target with increasing $n$ despite the enhanced QNG nature of both Fock states in the superposition~\cite{myatt2000decoherence}. Evaluating the quantum coherence aiming at the states $|\psi_{m,n}\rangle$ relies on the parameter $C_{m,n}$ defined by an extension of the definition in Eq.~(\ref{coherence}) for~$m>0$. Achieving the nonclassical coherence $C_{m,n}$ requires exceeding a threshold covering any coherence generated by the mixtures of classical coherent states of an oscillator. Similarly, the quantum non-Gaussian coherence and genuine $n$-phonon quantum non-Gaussian coherence certify insufficiency of the Gaussian evolution of the vacuum and the Gaussian evolution of the core state $|\widetilde{\psi}_n\rangle=\sum_{j=0}^{\max_{m,n}-1}c_j|j\rangle$, respectively, to overcome respective thresholds.

\begin{figure}[!t]
\begin{center}
\includegraphics[width=1\columnwidth]{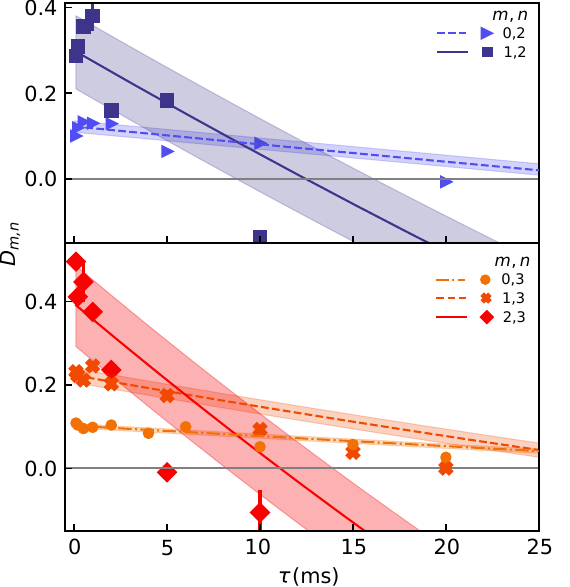}
\caption{Quantum non-Gaussian coherences for motional superpositions $|{\rm m,n}\rangle$. Red and blue data points correspond to evaluated measurement results for the binary Fock state superpositions with the higher Fock state $|2\rangle$ and $|3\rangle$, respectively.
The curves represent the upper limits on the achievable storage time due to the independently estimated amplitude damping through interaction with the thermal environment. The initial QNG depths of $|0,2\rangle$ and $|0,3\rangle$ are always smaller, but simultaneously the QNG depth is preserved longer than $|1,2\rangle$ and $|1,3\rangle$ (and even $|2,3\rangle$) states.}
\label{comparison}
\end{center}
\end{figure}

The coherences of superpositions of the number states of the form
\begin{equation}
|\psi_{m,n}\rangle := \frac{1}{\sqrt{2}}(|m\rangle + \exp(-i \phi_{\rm R_1})|n\rangle)
\end{equation}
with $\Delta_n=|m-n|=1;2$ of a trapped ion motion can be experimentally targeted by a modified Ramsey pulse sequence. It begins with the initial preparation of the electronic ground state and the motional Fock state $|g,m\rangle$ using a variation of the BSB and RSB resonant $\pi$-pulses~\cite{meekhof1996generation,mccormick2019quantum,podhora2022quantum}. Next, the electronic state superposition is implemented by carrier or BSB $\pi/2$ pulse for the target states with $\Delta=1$ and~2, respectively. This pulse also corresponds to the first pulse of the composite motional Ramsey interferometer. The following resonant Rabi interaction on the red motional sideband with the pulse area satisfying $\Omega \tau=2 \pi l$, where $l=\sqrt{\frac{1}{2}\frac{m}{n}(2 j+1)}\approx Z$, effectively maps the electronic coherence onto purely motional superposition for any integer $j$, which is optimized to be small to suppress decoherence during the state preparation~\cite{gardiner1997nonclassical}. It is followed by the Ramsey delay and the second - detection composite $\pi/2$ pulse. The observable defined in~(\ref{projector}) then acts in the motional $|m\rangle,|n\rangle$ basis.
A comparison of measured storage of binary superpositions states with various differences of energies $\Delta_n$ in Fig.~\ref{comparison} counter-intuitively signifies a clear preference of its larger values, which is evident despite the amplified susceptibility of the corresponding matrix elements $\rho_{m,n}$ to dephasing. Surprisingly, despite the lower initial QNG coherence, increasing $\Delta_n$ enhances the achievable storage time.

\section{Conclusions}

The presented criteria on quantum non-Gaussian coherences and the experimental verification of generated motional superpositions of a single trapped ion can be directly applied and compared to recent motional experiments in quantum electromechanics~\cite{chu2018creation,von2022parity}, photonic experiments in superconducting circuits~\cite{wang2017converting,wang2019heisenberg,reagor2016quantum,lu2021propagating,kudra2022experimental,kudra2022robust}, and quantum optics~\cite{yukawa2013generating,le2018slowing}. The achieved temporal storage of QNG coherences on the tens of millisecond timescales corresponds to more than~$10^4$ mechanical oscillation frequencies, which provides the feasibility of QNG mechanical memories~\cite{liu2023coherent,lake2021processing,fiore2011storing} and advances in processing on the oscillator-encoded qubits~\cite{hu2019quantum,fluhmann2019encoding} with the capability of deterministic quantum nonlinear interactions~\cite{roos2008nonlinear,ding2018quantum}. The developed QNG coherence criteria and depth are generally applicable, so a direct comparison of such different physical systems is possible. We explicitly compared the experimental and theoretical QNG depths presented in Fig.~\ref{fig:block_scheme}-c) and the depths of the superposition states in the stellar hierarchy~\cite{chabaud2021certification} that mixes both the diagonal and off-diagonal elements, see Supplementary Materials. Our approach allows for comparison of QNG coherences in the binary superpositions of the Fock states in quantum sensing~\cite{mccormick2019quantum}, quantum error correction~\cite{fluhmann2019encoding,hu2019quantum} and other emerging applications~\cite{burkhart2021error,maslennikov2019quantum,wang2020efficient,howl2021non,franke2023quantum}. Extension to ternary and more complex superpositions~\cite{hu2019quantum,fluhmann2019encoding} will clarify expected trade-offs between QNG coherences in the Fock-state basis. The single-mode quantum non-Gaussian coherences can be further extended to the generation and evaluation of multimode coherences also relevant in quantum sensing~\cite{zhang2018noon} and quantum error correction~\cite{chou2023demonstrating,gertler2021protecting}.

\begin{acknowledgments}
A.K., L.L., L.S. and R.F. acknowledge the support of the Czech Science Foundation under the project GA21-13265X and the H2020 European Programme under Grant Agreement 951737 NONGAUSS.
\end{acknowledgments}

\clearpage

\newpage

\setcounter{equation}{0}
\setcounter{figure}{0}
\setcounter{table}{0}
\setcounter{page}{1} \makeatletter
\renewcommand{\theequation}{S\arabic{equation}}
\renewcommand{\thefigure}{S\arabic{figure}}
\renewcommand{\theHequation}{Supplement.\theequation}
\renewcommand{\theHfigure}{Supplement.\thefigure}
\renewcommand{\bibnumfmt}[1]{[S#1]}
\renewcommand{\citenumfont}[1]{S#1}

\section*{Supplementary Material - Quantum non-Gaussian coherences of an oscillating atom}

\subsection{Stability of motional superpositions in a trapped ion}

The presented motional interferometric measurements crucially depend on the ability of coherent control of the ion’s spin-motional superpositions using laser pulses at 729 nm near-resonant with the $^{40}{\rm Ca}+$ quadrupole $4^{2}{\rm S}_{1/2}\leftrightarrow 3^2{\rm D}_{5/2}$ transition. The phase of each laser pulse is controlled with a precision of $0.02^\circ$ determined by the resolution of the employed direct digital synthesizers. The resulting radio frequency (RF) signal is applied to an acousto-optical modulator (AOM) driven with a center frequency of 552~MHz to provide 729~nm laser pulses in Ramsey interferometric sequences. High-frequency stability of this RF signal is achieved by the employment of a hydrogen maser clock for its digital synthesis, resulting in a short-term ($\sim 1$~s) frequency instability at the $10^{-13}$ level. This guarantees the repeatable realization of purely motional superpositions with a stable phase relation determined by the relative frequency and phase stability of the RF pulses generating in the AOM a consecutive train of optical pulses resonant with carrier, BSB, and RSB transitions. The long-term drifts of the laser frequency and magnitude of the applied magnetic field ($\sim 3.3$~Gauss) leave the repeatability of the phase of generated motional superposition unaffected. On the other hand, their short term frequency uncertainty manifested in a finite coherence time of the 729~nm laser interaction on the $4^{2}{\rm S}_{1/2}\leftrightarrow 3^2{\rm D}_{5/2}$ of 8~ms limits the achievable Ramsey contrasts through the decoherence of intermediate electronic state superpositions in the creation and observation composite motional Ramsey $\pi/2$ pulses.

The stability and repeatability of the generated motional superposition phase are dominantly limited by the slow drifts of electric DC potentials realizing the axial confinement applied through the two tip electrodes of the linear Paul trap, which manifest on the timescales of the measurement of full Ramsey fringes. The drift of the axial motional frequencies was estimated to~$0.44 \pm 0.11$~Hz/min using the Ramsey interferometry in the $|0,1\rangle$~state and observation of the drift of the set projection phase proportional to $\phi_\tau=\delta\omega \tau$, where $\delta\omega$ is the residual slowly varying difference between the actual motional frequency and the frequency of the precession of the generated superposition. This phase drift of motional superpositions affects the measurements only on very long timescales corresponding to long Ramsey delay times and long pulse sequences for states with a high $n$, where the total duration of the measurement approaches several minutes~\cite{mccormick2019quantum,turchette2000decoherence}. To suppress the effect of this long-term axial frequency drifts on the observable coherence $C$, the pulse sequence is repeated 25~times instead of 100~times for $n>2$, and the whole interference measurement is repeated four times to achieve the comparable statistical uncertainties of resulting coherence amplitudes. The smaller number of repetitions allows for sufficiently fast measurement runs compared to the long-term drifts of the measured Ramsey fringes and has no significant effect on the reliable estimation of coherences $C_{m,n}(t)$, as the measured interference fringes are still oversampled. The total duration of the measurement of a single data point with the $\tau\approx 0$ Ramsey delay in the takes about $300$~s for the states with small $\Delta_n=\{1,2\}$ and is reduced down to $100$~s for the measured for the presented largest $|0,n\rangle$ superpositions with $n=\{4,6\}$.

\subsection{Coherence of realized motional states}

\begin{figure}[t!]
\begin{center}
\includegraphics[width=1\columnwidth]{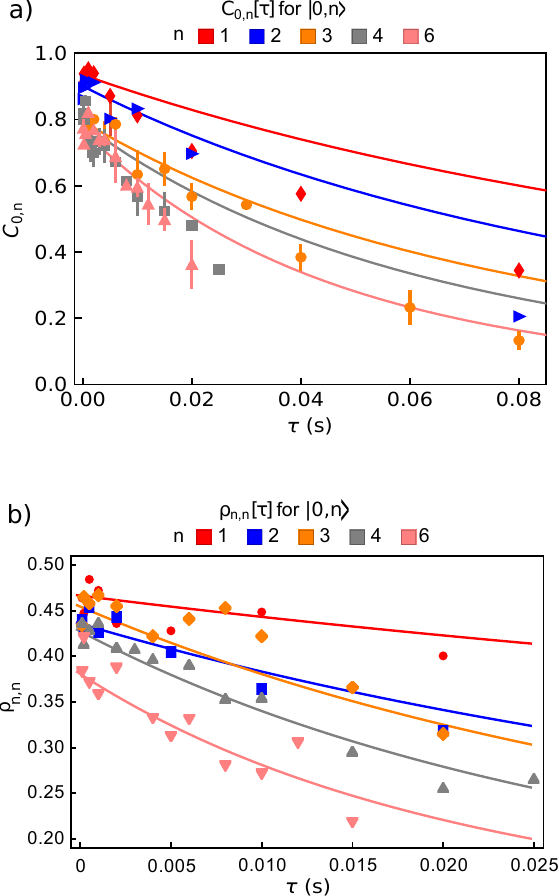}
\caption{a)~Measured coherences $C_{0,n}(\tau)$ for different Ramsey delay times on $|\psi_{0,n}\rangle$ states. b)~The corresponding diagonal matrix elements $\rho_{n,n}$ and their temporal decay. The solid curves represent in both graphs a model from~\cite{turchette2000decoherence} considering heating rate of an ion measured independently.}
\label{decayCoherences}
\end{center}
\end{figure}

Fig.~\ref{decayCoherences} presents the evaluated coherence $C_{0,n}(\tau)$ as a function of a Ramsey delay $\tau$ for $|\psi_{0,n}\rangle$ states. Differently to the QNG depth, the coherence decay accelerates for higher $n$, which is given by both enhanced thermalization of contributing $\langle n|\rho|n\rangle$ diagonal matrix elements~\cite{turchette2000decoherence} and also by increased sensitivity to different dephasing sources~\cite{milne2021quantum,McCormick2019}.
We validate that the upper limit on the loss of coherence by pure dephasing according to the eq.~(\ref{decayC}) is set by slow, but unavoidable heating of the motional state~\cite{brownnutt2015ion} by analysis of the temporal evolution of diagonal matrix elements in corresponding experimental settings. The phonon populations are reconstructed from measurements of the first BSB Rabi oscillations for the state emerging from the Ramsey generation $\pi/2$ pulse and following the time delay~$\tau$. The phonon number distribution $P(n)$ was estimated by fitting the resulting Rabi oscillations using the model~\cite{leibfried2003quantum}
\begin{equation}
P_{\rm g}(t) = \frac{1}{2}(1+\sum_{n}^{n_{\rm max}} P(n) \cos(\Omega_{\rm c} \eta \sqrt{n+1} t) \exp^{-\gamma(n)t}),
\label{eq:RabiOscillationFit}
\end{equation}
where $\Omega_{\rm c}=2 \pi\times 34.8\pm 0.1$~kHz is the Rabi frequency of the interaction on the corresponding carrier transition and $\eta = 0.063$ is the Lamb-Dicke parameter. The decay rate~$\gamma(n)$ accounts here mostly for the 729~nm laser intensity noise and beam pointing drifts. It scales with phonon number as~$\gamma(n) = (n+1)^x \gamma_0$, where $\gamma_0=2\pi\times (0.042\pm 0.005)$~kHz and~ $x=0.7$~\cite{leibfried2003quantum}, which we previously confirmed experimentally by measurement of the particular scaling of the Rabi oscillation decay with $n$ in this setup in the previous work focusing on the observation of genuine QNG properties of motional Fock states with a single trapped ion~\cite{Podhora2022}. The maximal number of phonons in~(\ref{eq:RabiOscillationFit}) was set such that $n_{\rm max}= n+2$, where $n$ is the larger number state from the generated binary superposition state $|m,n\rangle$. As illustrated in Fig.~\ref{decayCoherences}, the estimated populations confirm that their temporal evolution corresponds within experimental instabilities to a pure thermalization dynamics at the measured heating rate of~$3.2\pm 0.2$~phonons/s.


\subsection{Thresholds on quantum coherence}

Here, we detail the derivation of thresholds for the nonclassicality and quantum non-Gaussianity of coherence based on the measure defined as
\begin{equation}
    C_{m,n,\psi}(\rho)\equiv
    \frac{1}{2}\left\{\max_{\theta}\mbox{Tr}\left[S_{\theta,\psi}\rho\right]-\min_{\theta}\mbox{Tr}\left[S_{\theta,\psi}\rho\right]\right\},
\label{SM:cohDef}
\end{equation}
 where $S_{\theta,\psi}$ is an observable defined as
 \begin{equation}
     S_{\theta,\psi}=\cos \theta \left(\cos\psi X+\sin \psi Y\right)+\sin \theta Z
 \end{equation}
 with $X\equiv |m\rangle\langle n|+|n\rangle\langle m|$, $Y\equiv i\left(|m\rangle\langle n|-|n\rangle\langle m|\right)$ and $Z\equiv |m\rangle\langle m|-|n\rangle\langle n|$. The angles $\theta$ and $\psi$ represent longitude and latitude coordinates, respectively, on the Poincar\'{e} sphere with Fock states $|m\rangle$ and $|n\rangle$ at poles. Thus, the definition in Eq.~(\ref{SM:cohDef}) implies
 \begin{equation}
     C_{m,n}(\psi,\rho)=2|\cos \psi \rho_{m,n}|=2|\cos \psi \rho_{n,m}|.
 \end{equation}
 It shows that $C_{m,n}(\rho)$ depends only on the modul of off-diagonal elements of the density matrix and that the latitude $\psi$ modulates its value.


\subsubsection{Nonclassical coherence}

The nonclassical coherence certifies exceeding thresholds $C_{m,n}^{cl}$ corresponding to maximal $C_{m,n}$, defined by Eq.~(\ref{SM:cohDef}), that the classical states achieve. The maxima $C_{m,n}$ occur for pure classical states $|\alpha\rangle=D(\alpha)|0\rangle$ with $D(\alpha)\equiv \exp\left(\alpha a^{\dagger}+\alpha^* a\right)$ being the displacement operator and the thresholds $C_{m,n}^{cl}$ work out to be the maxima of the functions
\begin{equation}
    C_{m,n}(\alpha)=2\frac{|\alpha|^{m+n}}{\sqrt{m!n!}}e^{-|\alpha|^2}.
\end{equation}
Employing the sufficient condition for local maximum $\frac{d}{d|\alpha|}C_{m,n}=0$, we gain the thresholds
\begin{equation}
    C_{m,n}^{cl}=2\sqrt{\frac{(m+n)^{m+n}}{2^{m+n}}}e^{-\frac{m+n}{2}},
\end{equation}
which cover all coherence $C_{m,n}$ induced by classical states.

\subsubsection{Quantum non-Gaussian coherence}

The thresholds $C_{m,n}^{G}$ of the quantum non-Gaussian coherence are results of maximizing the coherence $C_{m,n}$ in Eq.~(\ref{SM:cohDef}) that the Gaussian states and their mixtures underlie. We can deduce that the thresholds $C_{m,n}^{G}$ correspond to maximal coherence achieved by pure Gaussian states $|\xi,\alpha \rangle\equiv D(\alpha)S(\xi)|0\rangle$, where $S(\xi)$ and $D(\alpha)$ are the squeezing operator and displacement operator acting on the vacuum. The amount of squeezing and amplitude of displacement are quantified by complex parameters $\xi$ and $\alpha$, respectively. Thus, maximizing $C_{m,n}$ corresponds to maximizing the related analytical functions over $\alpha$ and $\xi$. Further, we treat a complex number $z$ in the form $z=|z|e^{i\phi_z}$ of its magnitude and angle. The parametrization of the coherence measure achieved by any Gaussian state $|\gamma\rangle$ reads
\begin{equation}
\begin{aligned}
    &C_{m,n}(|\alpha,\xi \rangle)=\\
   & \frac{1}{\sqrt{n!m!}}|H_n\left(z\right)H_m\left(z^*\right)|\\
    &\times\frac{1}{\cosh{|\xi | }}\left(\frac{\tanh |\xi|}{2}\right)^{\frac{n+m}{2}}  e^{-|\alpha| ^2\left[ 1- \cos 2\phi \tanh{2\vert \xi  \vert}\right]},
    \label{SM:maxGC}
\end{aligned}
\end{equation}
where $\phi$ corresponds to a relative phase between phases of $\beta$ and $\xi$, $H_{n}(z)$ is the Hermite polynomial of the order $n$ and its complex argument $z$ reads
\begin{equation}
    z=\frac{|\alpha|e^{i\phi} }{\sqrt{\sinh 2|\xi|}}.
    \label{SM:zDef}
\end{equation}
Since any complex number $z$ satisfies $\max_{\theta}z e^{i\theta}+z^*e^{-i\theta}=|z|$ and $\min{\theta}z e^{i\theta}+z^*e^{-i\theta}=-|z|$, we can identify the respective quantifier of coherence
\begin{equation}
\begin{aligned}
   & C_{m,n}(|\xi,\alpha \rangle)=\frac{1}{2\sqrt{n!m!}\cosh{|\xi | }}\left(\frac{\tanh |\xi|}{2}\right)^{\frac{n+m}{2}}\\
    &\times |H_n\left(z\right)H_m\left(z^*\right)|e^{-|\alpha| ^2\left[ 1- \cos 2\phi \tanh{2\vert \xi  \vert}\right]}.
    \label{SM:minGC}
\end{aligned}
\end{equation}
Thus, the maximizing can be solved numerically for each particular choice of $m$ and $n$. To speed up the numerical maximizing, we can reveal an analytical constraint among the parameters. It stems from the equation $\nabla C_{m,n}=0$. Let $F_{m,n}(z,z^*)$ denote $\max_{\theta}\left[H_m(z)H_n(z^*)e^{i\theta}+c.c.\right]$ with $z$ being defined in (\ref{SM:zDef}). This allows us to express the differentiation as
\begin{equation}
\begin{aligned}
    &\partial_{|\alpha|} C_{m,n}=\frac{1}{\sqrt{m!n!}\cosh 2 |\xi|}\left(\frac{\tanh 2|\xi|}{2}\right)^{\frac{n+m}{2}} \\
    &\times e^{-|\alpha| ^2(1-\cos 2\phi \tanh |\xi |)}\left[\partial_{|\alpha|}F_{m,n}(z,z^*)\right.\\
    &\left. -2|\alpha| (1-\cos 2\phi \tanh |\xi |)F_{m,n}(z,z^*)\right]=0\\
    &\partial_{|\xi|} C_{m,n}=\frac{1}{\sqrt{m!n!}\cosh 2 |\xi|}\left(2\tanh{|\xi|}\right)^{\frac{n+m}{2}} \\
    &\times e^{-|\alpha| ^2(1-\cos 2\phi \tanh |\xi |)}\Big\{ \partial_{|\xi|}F_{m,n}(z,z^*)\\
    & -\left[2\tanh{|\xi|}+\frac{2}{\sinh 2|\xi|}(1+m+n+2|\alpha|^2 \cos 2\phi \tanh |\xi |)\right]\\
    &\times F_{m,n}(z,z^*)\Big\}=0
\end{aligned}
\label{eqsAlf}
\end{equation}
Note, the last equation corresponding to differentiation with respect to $\phi$ is irrelevant here. Since the rules of differentiation composite functions guarantee
\begin{equation}
\begin{aligned}
    |\alpha| \partial_{|\alpha|}&F_{m,n}(z,z^*)=\\
    &-\tanh{2|\xi|}\partial_{|\xi|}F_{m,n}(z,z^*),
\end{aligned}
\end{equation}
equations (\ref{eqsAlf}) implies
\begin{equation}
    \begin{aligned}
        &|\alpha|dF_{n,m}(z,z^*) \cosh 2|\xi|-F_{n,m}(z,z^*)\sqrt{\sinh 2|\xi|}\\
        &\times \left(m+n-2\sinh^2 |\xi|+2|\alpha|^2 \cos 2\phi \tanh |\xi|\right)=0\\
        & dF_{m,n}(z,z^*)+2|\alpha|F_{m,n}(z,z^*)(-1+\cos 2\phi \tanh |\xi|=0,
    \end{aligned}
\end{equation}
where $dF_{m,n}(z,z^*)\equiv\partial_{|\alpha|}F_{m,n}(z,z^*)$. These equations can be satisfied for some positive $F_{m,n}(z,z^*)$ only if
\begin{equation}
    |\alpha|^2=\frac{(1+m+n)\mbox{sech\ } 2|\xi|-1}{2(1-\cos 2\phi \tanh{2 |\xi|})}.
    \label{betaRel}
\end{equation}
This simplifies the task to maximizing over two parameters $\phi$ and $|\xi|$. We confirmed this maximizing always leads to $\phi=0$ up to $\max \left(m,n\right) \leq 10$.

\subsubsection{Genuine $n$-phonon quantum non-Gaussian coherence}

The determination of the thresholds exposing the genuine $n$-phonon quantum non-Gaussianity relies on the numerical maximizing of extensive functions that represent the coherence quantifier achieved by the Gaussian dynamics of the core state $|\widetilde{\psi}_{n-1}\rangle=\sum_{k=0}^{n-1}c_k|k\rangle$.
To express them, we introduce an auxiliary function
\begin{equation}
\begin{aligned}
    S_{m,n}&(x,y,z)=\\
    &\sum_{i=0}^{\min \left[ m,n\right]}\frac{\sqrt{n! m!}}{i!(m-i)!(n-i)!}z^{m-i}H_{m-i}(y)H_{n-i}(x),
\end{aligned}
\end{equation}
where $H_n$ is the Hermite polynomial of the order $n$, and the amplitude $a_{m,n}(\xi,\alpha)=\langle m|S(\xi)D(\alpha)|n\rangle$, yielding \cite{Kral1990}
\begin{equation}
\begin{aligned}
   & a_{m,n}(\xi,\alpha)=\left(\frac{1}{m!n! \cosh \mu}\right)^{1/2}\left(\frac{\tanh |\xi|}{2}\right)^{n/2}\\
   &\left(\frac{4}{\sinh 2|\xi|}\right)^{m/2} e^{-\frac{1}{2}\left(|\beta|^2-\beta^2\tanh |\xi |\right)}\\
   &\times S_{m,n}\left(\frac{\beta}{\sqrt{\sinh 2|\xi|}},i\frac{\alpha^*}{\sqrt{\sinh 2|\xi|}},i\frac{\sinh |\xi|}{\sqrt{2}}\right)
\end{aligned}
    \label{amn}
\end{equation}
with $\alpha=\beta \cosh |\xi|-\beta^* \sinh |\xi|$. Further, we define an $n\times n$ matrix $\boldsymbol{\mathcal{G}}(\theta,\alpha,\xi)$ by its elements
\begin{equation}
    \mathcal{G}_{i,j}(\theta,\beta,\xi)=a_{i,m}(\xi,\alpha)a_{j,n}^*(\xi,\alpha) e^{i\theta}+c.c.
\end{equation}
All together, the coherence quantifier of the states $|\gamma,\widetilde{\psi}_{n-1}\rangle\equiv S(\xi)D(\alpha)|\widetilde{\psi}_{n-1}\rangle$ reads
\begin{equation}
    C_{m,n}(|\gamma,\widetilde{\psi}_{n-1}\rangle)=2\max_{\theta}\textbf{c}^*\cdot \boldsymbol{\mathcal{G}}(\theta,\xi,\alpha) \cdot \textbf{c}^T,
\end{equation}
where $\textbf{c}=(c_0,...,c_{n-1})$ is a vector determining the amplitudes $c_k$ in the core state $|\widetilde{\psi}_{n-1}\rangle=\sum_{k=0}^{n-1}c_k |k\rangle$.

The thresholds $C_{m,n}^{G_n}$ correspond to the maxima over both all the Gaussian evolution $S(\xi)D(\alpha)$ and all the vectors $\textbf{c}$. The later maximizing over $\textbf{c}$ works out to be the maximal eigenvalue of the matrix, i.e this implies \cite{Fiurasek2022}
\begin{equation}
    C_{m,n}^{G_n}=2\max_{\theta,\xi,\alpha}\lambda(\theta,\xi,\alpha),
\end{equation}
where $\lambda(\theta,\beta,\xi)=\max \mbox{eig } \boldsymbol{\mathcal{G}}(\theta,\beta,\xi)$ with $\mbox{'eig'}$ denoting the vector of the eigenvalues. The maximizing over $\theta$, $\xi$ and $\alpha$ can be performed only numerically.

\subsubsection{Intrinsic quantum non-Gaussian coherences}

In addition, we compared the coherences $C_{0,n}$ to thresholds achieved by any Gaussian transformation of arbitrary Fock state $|m\rangle$. It does not consider any superposition of Fock states up to $n-1$ under Gaussian transformation. Therefore, we only maximize $C_{0,n}(|\alpha,\xi,m\rangle)$ over the states $|\alpha,\xi,m\rangle\equiv S(\xi)\textsf{D}(\alpha)|m\rangle$ with $S(\xi)$ being the squeezing operator and $\textsf{D}(\alpha)$ is displacement operator. We do numerical optimization over all $\alpha$, $\xi$ and also $m$. We can establish that the values $0.93$, $0.70$, $0.63$ and $0.55$ represent consecutively the maximal $C_{0,n}(|\alpha,\xi,m\rangle)$ with $n=1,...,4$ for this hierarchy.
For $n=1$, the thresholds coincide since the optimal Fock state is $m=0$.  Therefore, the superposition of Fock states up to $n-1$ is essential to construct the genuine hierarchy of quantum non-Gaussian coherence.

\subsection{Dephasing depth of quantum non-Gaussian coherence}
\label{sec:depth}

Exceeding a threshold can be operationally quantified in terms of the sensitivity to pure dephasing. A dephasing channel deteriorates the state $|\psi_{m,n}\rangle=(|m\rangle+\exp(i\phi)|n\rangle)/\sqrt{2}$ by random changes of the phase $\phi$ with a variance $\Gamma=\langle\phi^2\rangle$. Thus, the dephasing channel reduces the off-diagonal elements $\rho_{m,n}=\langle m|\rho|n\rangle$ of a state $\rho$ according to $\rho_{m,n}\rightarrow \rho_{m,n} \exp\left[-\Gamma/2 (m-n)^2\right]$. The depth $D_{m,n}$ of the quantum non-Gaussian coherence corresponds to the maximal $\Gamma$ preserving the quantum non-Gaussian coherence.

\subsection{Nonclassical and quantum non-Gaussian coherences between a pair of Fock states}

The quantum coherence between a pair of the different Fock states $|m\rangle$ and $|n\rangle$ can exhibit various forms in the superposition  $|\sigma_{m,n}\rangle=\alpha |n\rangle+\beta |m\rangle$. Measurement needed to detect such quantum coherence employs the observables $X_{m,n}=|n\rangle \langle m|+|m\rangle \langle n|$ and $Y_{m,n}=i\left(|n\rangle \langle m|-|m\rangle \langle n|\right)$. We can also consider the operator $Z_{m,n}=|m\rangle \langle m|-|n\rangle \langle n|$ to complete all the measurement on the Poincar\' e sphere with the Fock states $|m\rangle$ and $|n\rangle$ on its poles \cite{Bloch1946}. Therefore, we can always find a observable
\begin{equation}
    S_{\theta,\psi}=\cos \psi \left(\cos \theta X+\sin \theta Y\right)+\sin \psi Z,
\end{equation}
on a vector in this Poincar\'e sphere, which obeys $S_{\theta,\phi} |\sigma_{m,n}\rangle=|\sigma_{m,n}\rangle$ for some specific angles $\psi$ and $\theta$ associated with the coefficient $\alpha$ and $\beta$ in the state $|q_{m,n}\rangle$. Using an analogy with classical interferometry,  angle $\theta$ represents the phase control to observe an interference fringe and $\psi$ is set such that it maximizes the visibility \cite{Loudon2000}.
Based on this classical analogy, we introduce a quantifier of coherence according to
\begin{equation}
    C_{m,n,\psi}(\rho)=\frac{1}{2}\lbrace \max_{\theta}\mbox{Tr}\left[S_{\theta,\psi} \rho\right]-\min_{\theta}\mbox{Tr}\left[S_{\theta,\psi}\rho \right]\rbrace,
    \label{defCnm}
\end{equation}
which identifies the difference between maximum and minimum of the measurement $S_{\theta,\phi}$ over phase $\theta$, and therefore the coherence quantifier (\ref{defCnm}) does not depend on $\langle Z \rangle$. Also, $C_{m,n,\theta}=\cos \theta C_{m,n,0}$ holds identically for all states, and therefore we set $\psi=0$ without losing the generality and use the notation $C_{m,n}=C_{m,n,0}$ in the following. Unlike in the classical interferometry, we do not normalise $C_{m,n}$ to keep it convex in the state $\rho$. The definition (\ref{defCnm}) implies the coherence quantifier $C_{m,n}(\rho)$ is given by the off-diagonal element according to $C_{m,n}(\rho)=2|\langle m|\rho|n\rangle|$. Also, $C_{m,n}(\rho)$ guarantees the convexity in a density matrix, i.e. any $\rho$ expressed as $\rho=p_1 \rho_1+p_2 \rho_2$, where $\rho_1$ and $\rho_2$ correspond to some density matrices, obeys
\begin{equation}
    C_{m,n}(\rho) \leq p_1 C_{m,n}(\rho_1)+p_2 C_{m,n}(\rho_2)
    \label{Meth:convexity}
\end{equation}
due to the triangle inequality. Thus, mixing states does not increase the coherence quantifier.

The coherence in Fock state basis is already inherent to coherent states $|\alpha\rangle$ introduced in optics \cite{Glauber1963a,Sudarshan1963}. Their coherence quantifier is, however, covered by the threshold
\begin{equation}
    C_{m,n}^{cl}\equiv \frac{2}{\sqrt{m!n!}}\left(\frac{m+n}{2}\right)^{\frac{n+m}{2}}e^{-\frac{m+n}{2}}.
\end{equation}
To identify even higher threshold, we allow for states $|\alpha,\xi\rangle=S(\xi)D(\alpha)|0\rangle$, resulting from combined action of the squeezing operator $S(\xi)$ and the displacement operator $D(\alpha)=\exp \left(\alpha^* a-\alpha a^{\dagger}\right)$ \cite{Yuen1976two}.
Apart these cases, the respective thresholds $C^{\rm G, min}_{m,n}$ follow as:
\begin{equation}
    \begin{aligned}
        C^{\rm G, min}_{m,n}&=\max_{|\xi|}\frac{2}{\sqrt{n!m!}\cosh |\xi|}|H_n(z(|\xi|))H_m(z(|\xi|))|\\
        &\times e^{-\frac{1}{2}\left[1+m+n-\cosh(2|\xi|)\right]\left[ 1+\tanh{\vert \xi  \vert}\right]}
    \end{aligned}
    \label{CnmMax}
\end{equation}
where $H_n$ is the Hermite polynomial of the order $n$ and
\begin{equation}
   z(|\xi|)=\frac{1+m+n-\cosh{2|\xi|}}{(1-\tanh{2|\xi|)\sinh{4|\xi|}} }.
\end{equation}
To increase demands imposed on the coherence quantifier, we explore $C_{m,n}$ (with $m<n$) induced by the Gaussian dynamics $S(\xi)D(\alpha)$ affecting the core states $|\widetilde{\psi}_{n-1}\rangle=\sum_{l=0}^{n-1}c_l|l\rangle$ instead of the vacuum \cite{Lachman2019,Chabaud2020}. The state $|\widetilde{\psi}_{n-1}\rangle$ alone yields $C_{m,n}=0$, and therefore it needs to be evolved by the Gaussian dynamics to exhibit $C_{m,n}>0$. The respective thresholds $C_{m,n}^{G_n}$ covering these states in $C_{m,n}$ can be derived numerically~\cite{Fiurasek2022}. We verified values of all the derived thresholds by Monte-Carlo simulations, confirming both validity and also presence of the Gaussian states near thresholds.
\begin{figure}[!t]
\begin{center}
\includegraphics[width=1\columnwidth]{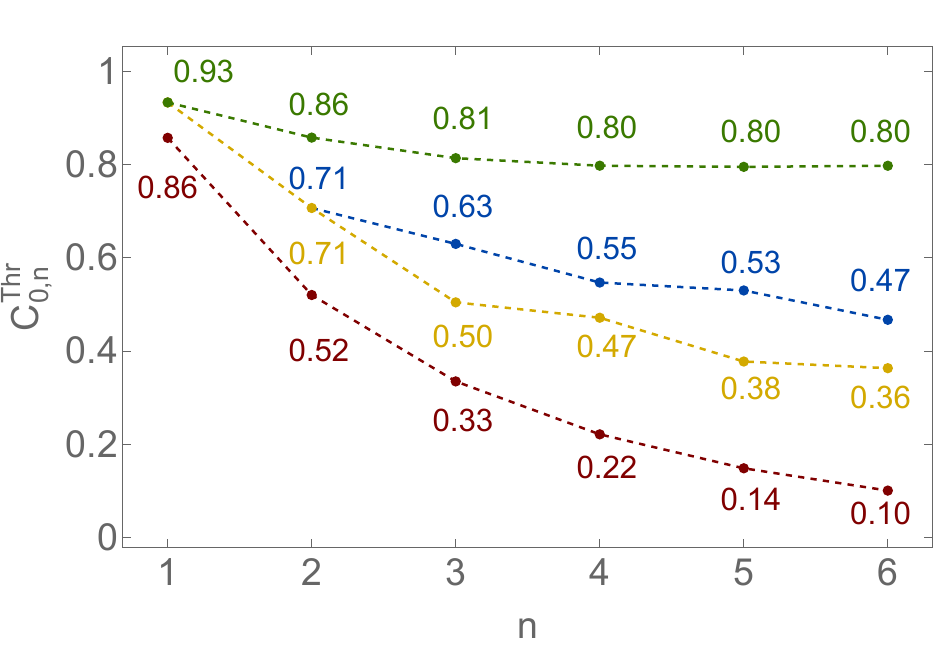}
\caption{The figure presents numerically derived values of thresholds $C_{0,n}^{\rm cl}$ of nonclassical coherence (brown), $C_{0,n}^{G}$ of minimal quantum non-Gaussian coherence (yellow), intrinsically quantum non-Gaussian coherence $C_{0,n}^{\rm G, int}$ (blue), and hierarchy of genuine $n$-phonon quantum non-Gaussian coherence $C_{0,n}^{\rm G_n}$ (green).
}
\label{fig:thresAll}
\end{center}
\end{figure}

Fig.~\ref{fig:thresAll} depicts these thresholds and compares them with the thresholds covering only the Gaussian dynamics of the vacuum. Thus, we certify nonclassical coherence, quantum non-Gaussian coherence, intrinsically quantum non-Gaussian coherence, or genuine $n$-phonon quantum non-Gaussian coherence by measuring $C_{m,n}$ exceeding the threshold $C_{m,n}^{\rm cl}$, $C_{m,n}^{\rm G,min}$, $C_{m,n}^{\rm G, int}$, or $C_{m,n}^{{\rm G}_n}$, respectively. Interestingly, the hierarchy thresholds $C_{0,n}^{G_{n}}$ numerically saturates at the value approximately~0.8.

To complete the analysis and information about our experiment, we provide the estimated fidelities of the realized superposition states $|\psi_{0,n}\rangle$ and corresponding thresholds on fidelities required for validation of the genuine QNG coherent aspects $F_{0,n}^{\rm G_n, Thr}$, as summarized in Table~\ref{table}. The state fidelities combine diagonal and off-diagonal elements (coherences); therefore, thresholds are generally different than for coherences alone. The level of genuine quantum non-Gaussian superposition states was achieved for $n=2,3,4$. The depths of the QNG coherences are systematically more sensitive to dephasing than the depths of QNG fidelities influenced by the diagonal elements. In addition, evaluating coherences solely from the off-diagonal elements provides more direct information on their intrinsic QNG coherent aspects.

\begin{widetext}

\begin{table}[hb!]
\begin{tabular}{|c|c|c|c|c|c|}
\hline
$n$        & 1       & 2     & 3 &  4 &   6 \\ \hline \hline
$F_{\rm Exp}$ & $0.92\pm 0.07$ &  $0.90\pm 0.04$ & $0.86\pm 0.03$ & $0.88\pm 0.03$ & $0.77\pm 0.06$ \\ \hline
$F_{0,n}^{\rm G_n, Thr}$ & 0.95 & 0.82 & 0.84 & 0.79 & 0.82  \\ \hline
$D_{0,n}^{\rm Fid, Ideal}$ & 0.23	& 0.23 & 0.09 & 0.07 & 0.03 \\ \hline
$D_{0,n}^{\rm Fid, Exp}$ & 0	& 0.14 & 0.06 & 0.03 & 0 \\ \hline \hline

$C_{0,n}^{\rm Exp}$ & $0.95 \pm 0.01$ & $0.917 \pm 0.004$ & $0.81 \pm 0.03 $ & $0.84 \pm 0.04$ & $0.80 \pm 0.05$\\ \hline
$C_{0,n}^{\rm G_n, Thr}$ & 0.93	& 0.86	& 0.81 &	0.80 & 0.80 \\ \hline
$D_{0,n}^{\rm Coh, Ideal}$ & 0.14 & 0.08 & 0.05 & 0.03 & 0.01 \\ \hline
$D_{0,n}^{\rm Coh, Exp}$ & 0.03 & 0.03 & 0 & 0.01 & 0 \\ \hline
\end{tabular}
\caption{ Summary of the realized state fidelities ($F_{\rm Exp}$) with ideal binary superposition states $(|0\rangle+|n\rangle)/\sqrt{2}$, their corresponding thresholds on the genuine QNG superposition states ($F_{0,n}^{\rm G_n, Thr}$) according to the Ref.~\cite{Chabaud2021}, and depths under the pure dephasing process $D_{0,n}^{\rm Fid}$, as defined in the Secion~\ref{sec:depth}. In order to allow for a direct comparison, we list also the explicit values of the measured coherences $C_{0,n}^{\rm Exp}$, corresponding thresholds on the genuine $n$-phonon quantum non-Gaussian coherences $C_{0,n}^{\rm G_n, Thr}$, and dephasing depths $D_{0,n}^{\rm Coh}$.}
\label{table}
\end{table}

\end{widetext}

In some cases, we can employ the Gaussian dynamics of the Fock states to achieve nonclassical or quantum non-Gaussian coherence although the Fock states themselves do not exhibit any of these properties. For example, the state $D(\alpha)|1\rangle$ gains $C_{0,2}=0.652>C_{0,2}^{\rm cl}$ for $|\alpha|^2=0.586$, which implies the nonclassical coherence. Similarly, we can establish that the state $S(\xi)D(\alpha)|1\rangle$ with $|\alpha|^2=0.937$ and $\xi=0.2$ exhibits $C_{0,2}$ below $C_{0,2}^{\rm G,min}$, but $C_{0,3}=0.6293>C_{0,3}^{\rm G,min}$, which signifies the quantum non-Gaussian coherence. It demonstrates that Gaussian evolution can induce the quantum non-Gaussian coherence from Fock states. In contrast, the genuine $n$-phonon quantum non-Gaussian coherence detected by the hierarchy cannot be produced by Gaussian dynamics of any Fock state \sout{(see Supplementary Materials)}.

\end{document}